\documentclass[preprint,12pt]{elsarticle}

\usepackage{amssymb}
\usepackage{siunitx}
\usepackage{booktabs}

\journal{Computerized Medical Imaging and Graphics}

\begin{document}

\begin{frontmatter}



\title{Predicting 4D Liver MRI for MR-guided Interventions}

\author[inst1]{Gino Gulamhussene\corref{cor1}}
\author[inst1]{Anneke Meyer \fnref{ref1}}
\author[inst1]{Marko Rak \fnref{ref1}}
\author[inst1]{Oleksii Bashkanov}
\author[inst2]{Jazan Omari}
\author[inst2]{Maciej Pech}
\author[inst1]{Christian Hansen}
\cortext[cor1]{gino.gulamhussene@ovgu.de}
\fntext[ref1]{These authors contributed equally.}

\affiliation[inst1]{organization={Otto-von-Guericke University, Faculty of Computer Science},
            addressline={Universitätsplatz 2}, 
            city={Magdeburg},
            postcode={39106}, 
            state={Saxony-Anhalt},
            country={Germany}}

\affiliation[inst2]{organization={University Hospital Magdeburg, Department of Radiology and Nuclear Medicine},
            addressline={Leipziger Straße 44}, 
            city={Magdeburg},
            postcode={39120}, 
            state={Saxony-Anhalt},
            country={Germany}}

\begin{abstract}
Organ motion poses an unresolved challenge in image-guided interventions. In the pursuit of solving this problem, the research field of time-resolved volumetric magnetic resonance imaging (4D MRI) has evolved. However, current techniques are unsuitable for most interventional settings because they lack sufficient temporal and/or spatial resolution or have long acquisition times. In this work, we propose a novel approach for real-time, high-resolution 4D MRI with large fields of view for MR-guided interventions. To this end, we trained a convolutional neural network (CNN) end-to-end to predict a 3D liver MRI that correctly predicts the liver's respiratory state from a live 2D navigator MRI of a subject. Our method can be used in two ways: First, it can reconstruct near real-time 4D MRI with high quality and high resolution ($209\times128\times128$ matrix size with isotropic $\SI{1.8}{\milli\meter}$ voxel size and $\SI{0.6}{\second\per volume}$) given a dynamic interventional 2D navigator slice for guidance during an intervention.
Second, it can be used for retrospective 4D reconstruction with a temporal resolution of below $\SI{0.2}{\second\per volume}$ for motion analysis and use in radiation therapy. We report a mean target registration error (TRE) of \num{1.19\pm 0.74}\si{\mm}, which is below voxel size. We compare our results with a state-of-the-art retrospective 4D MRI reconstruction. Visual evaluation shows comparable quality. We show that small training sizes with short acquisition times down to $\SI{2}{\min}$ can already achieve promising results and $\SI{24}{\min}$ are sufficient for high quality results. Because our method can be readily combined with earlier methods, acquisition time can be further decreased while also limiting quality loss. We show that an end-to-end, deep learning formulation is highly promising for 4D MRI reconstruction. 
\end{abstract}


\begin{highlights}
\item 4D Liver MRI prediction from dynamic fixed-position 2D navigators
\item Fast high-resolution 4D MRI method ($\SI{0.6}{\second\per volume}$) showing variable respiratory motion 
\item End-to-end deep-learning formulation and image based regression
\item Reduction of pre-interventional data acquisition time (2 minutes training data)
\item Publication of MRI data from 20 subjects as training data for deep learning
\end{highlights}


\end{frontmatter}

\section{Introduction}
\label{sec:intro}
4D MRI is not easily available from a medical point of view. Part of this is due to the significant amount of data needed to reconstruct different breathing states. Development of 4D MRI methods is becoming more advanced and could soon make 4D MRI readily available for use in clinical scenarios like needle guidance during cancer intervention on a liver.  
We can classify most of the proposed 4D MRI methods for large fields of view (FOVs) and high temporal resolution along two lines. The first line differentiates whether sequence programming and altered k-space sampling is used, or whether readily available standard sequences are used. The second line differentiates whether a "representative", i.e., single breathing cycle or any sequence of breathing patterns, is reconstructed. Virtually all proposed methods - either prospective or retrospective - are not able to acquire or reconstruct a 4D MRI in real-time. In this work we focus on the reconstruction of arbitrary sequences of breathing patterns in retrospection using readily available clinical sequences. Furthermore, we show that our method is capable of predicting real-time 4D MRI utilizing a 2D live navigator slice. 

In 2007, von Siebenthal et al. \cite{siebenthal2007} proposed a 4D MRI reconstruction method for arbitrary breathing sequences. They proposed to retrospectively sort 2D cine, i.e., dynamic, MRI sequences of interleaved navigator and data slices based on the navigator's breathing state. Building upon that work, further methods were proposed to improve image quality. 
To this end, Tanner et al. \cite{tanner2014} predicted unseen data slices for navigator slices.
Gulamhussene et al. \cite{gulamhussene2020} improved reconstruction speed and robustness against out-of-plane motion in the navigator by applying template updates. 
Other work focused substantial effort on the reduction of acquisition time.
In this respect, Celicanin et al. \cite{celicanin2015} simultaneously acquired navigator and data slices, cutting the acquisition time in half. 
Karani et al. \cite{karani2018} predicted every second navigator using a convolutional neural network (CNN), shortening acquisition time by one fourth. 
Zhang et al. \cite{zhang2018} also performed interpolation of navigators by utilizing an intermediate motion field prediction using a CNN.
Navest et al. \cite{navest2020} used information always present in the raw MRI readout, eliminating the need for acquisition of an actual navigator frame.

In contrast, other approaches propose to reconstruct one "representative" breathing cycle \cite{cai2011, tryggestad2013, paganelli2015, van2018, romaguera2019, meschini2019, yang2020}.
The notion of single representative breathing cycles is common in radiation therapy planning and execution. 
Another group of 4D MRI techniques used sequence programming and/or altered k-space sampling \cite{hu2013, deng2016, han2017, qiu2019}. All of these reconstructed average, i.e., representative, breathing patterns as well. 
Only in 2019 did Yuan et al. \cite{yuan2019} propose the first near real-time large FOV 4D MRI reconstruction technique. They used sequence programming, attaining a high temporal resolution ($615 \si{\milli\s}$) with moderate spatial resolution ($128\times128\times56$ matrix size, $2.7\times2.7\times4.0 \si{\mm}^3$ voxels).

In this work, we propose a novel approach that uses a CNN to solve two distinct problems: First, it can be used to predict 3D full-liver MRIs with high spatial resolution ($209\times128\times128$ matrix size, $1.8\times1.8\times1.8 \si{\mm}^3$ voxels) live and in near real-time ($600 \si{\ms}$). Second, it can be used as a retrospective method to reconstruct 4D MRIs with high temporal resolution ($116 \si{\ms}$) and the same spatial resolution for arbitrary physiological breathing patterns extracted from 2D navigator sequences. Our method is capable of reconstructing 3D liver MRIs even with drastically reduced training data, cutting acquisition times to only a few minutes compared to onboard 4D MRI techniques. 

\section{Material and methods}

\subsection{Training data}
\label{sec:training_data}

We published part of the data used in this work in a data repository previously \cite{gulamhussene2019}. We acquired further data of 7 new subjects and made it also publicly available for this work in a data repository \cite{gulamhussene2021}. The data contains anonymised DICOM images as well as a detailed MRI acquisition protocol.  The data was acquired on a MAGNETOM Skyra MRI scanner (Siemens Medical Solutions, Erlangen, Germany). In total, the data comprises MRIs from 20 healthy volunteers. For each subject, a static 3D liver MRI, a 2D navigator MRI reference sequence, and several interleaved MRI sequences were acquired. The latter are sequences with alternating 2D navigator and 2D data slices, also called image slices. 3D volumes and 2D slices of the same subject share common scanner coordinates. The volume was acquired with a STAR VIBE sequence (320$\times$320$\times$72-88 matrix size, \SI{3}{\mm} slice thickness, $1.19\times\SI{1.19}{\mm}^2$ voxels, 0\% phase oversampling, 44.4\% slice oversampling, \SI{380}{\mm} FOV read, 100\% FOV phase, \SI{2.83}{\mm} TR, \SI{1.48}{\ms} TE, \ang{9} flip angle, 7/8 slice partial Fourier). A STAR VIBE sequence allows for the imaging of a still 3D MRI and is based off of an actually moving target, such as the liver during free breathing.
 
Navigator and data slices were acquired with a TRUFI sequence (\SI{39.96}{\ms} TR, \SI{3.33}{\ms} echo spacing, \SI{1.49}{\ms} TE, \ang{30} flip angle, \SI{676}{\Hz\per voxel} readout bandwidth, 176 $k_x$ base resolution, 80\% phase resolution, 14$\times$176 matrix size, $1.8\times\SI{1.8}{\mm}^2$ in-plane resolution, \SI{4}{\mm} out of plane resolution, $255\times\SI{320}{mm}^2$ FOV). For faster measurement, a partial Fourier was used sampling 5/8 of the k-space asymmetrically in phase-encoding direction, i.e., roughly 60\% of the $k_y$ lines, resulting in 88 actually acquired lines.
The acquisition takes place with high speed (\SI{166}{\ms\per slice}), with contrast just good enough to detect the respiratory motion. No body array coil was used, limiting acquisition to the bore's fixed receiver coil.
 
\begin{figure}[t]
    \begin{center}
        \includegraphics[width=.85\textwidth]{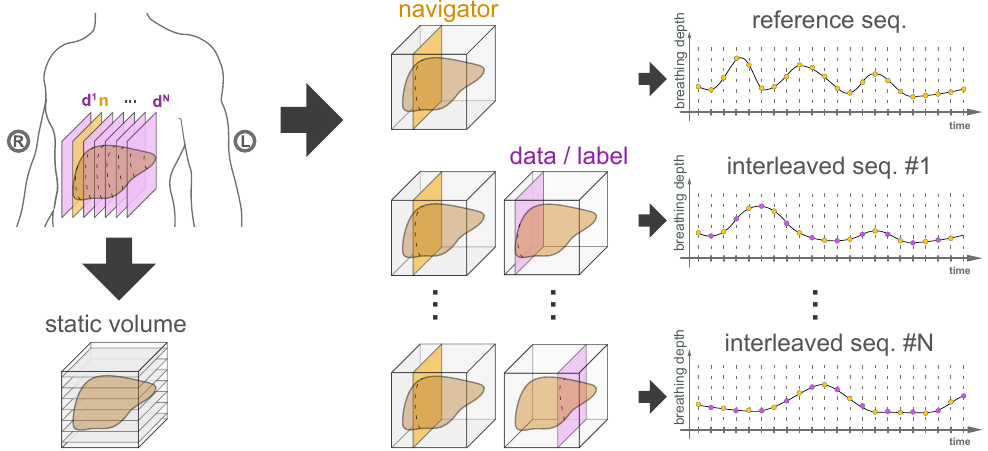}
    \end{center}
    \caption{For each subject a static 3D volume was acquired with a STAR VIBE sequence (seq.) with axial slice orientation. A reference sequence comprised only of navigator frames was acquired using a TRUFI sequence and several interleaved sequences were acquired using the same TRUFI sequence, equidistantly sampling the liver in sagittal orientation.}
    \label{fig:teaser}
\end{figure}

Reference and interleaved sequences will be described in the following. 
A reference sequence is a 2D cine MRI sequence of so-called navigator frames, where a slice is acquired every 166 ms. A schematic depiction can be found in Fig. \ref{fig:teaser}. Navigator frames represent an image plane, in which the respiratory motion is visible via vessel cross-sections. In our case, the navigator is a sagittal slice that intersects the liver. This sequence is our reference for 4D reconstruction. It contains a natural succession of different breathing cycles and patterns, like shallow or deep, thoracic or abdominal breathing, and is thus physiologically meaningful. In our case, the reference sequence always comprises 513 time points, covering 85 seconds (typically about 20 breathing cycles). 

Each interleaved sequence (see Fig. \ref{fig:teaser}) consists of 300 to 400 data and navigator slices (\SI{166}{\ms} per slice). It can be thought of as a series of pairs of these navigator and data slices. The navigator slices will be part of the networks training input, and the data slices will be the training label images. The training will be discussed in Sec. \ref{sec:training}. The navigator slices are positioned exactly as they are in the reference sequences. Each interleaved sequence is imaged for one minute before moving the imaging plane of the data slice \SI{4}{\mm} to the left, while keeping the navigator position fixed. This is repeated until the entire liver is covered. This enables the CNN to learn spatial relations between navigator and data frames to later reconstruct the whole liver volume from a single navigator slice. The total number of training sequences per subject ranges between 38 and 57, depending on the size of the subjects’ liver. Thus, the overall acquisition time for a subject ranged between 40 and \SI{80}{\min}, excluding the time needed for imaging localizers, determining the navigator position and setting up the interleaved sequences. The latter averages to about \SI{15}{\min} per subject.
The Otto-von-Guericke-University Magdeburg ethics board approves our study “Studies with healthy subjects in 3 Tesla for methodological development of MRI experiments” (approval number 172/12), stating they concluded that there are no ethical concerns and that this approving assessment is made based on unchanged conditions. Oral and written consent was obtained during the study.

\begin{figure}[ht]
    \begin{center}
        \includegraphics[width=\textwidth]{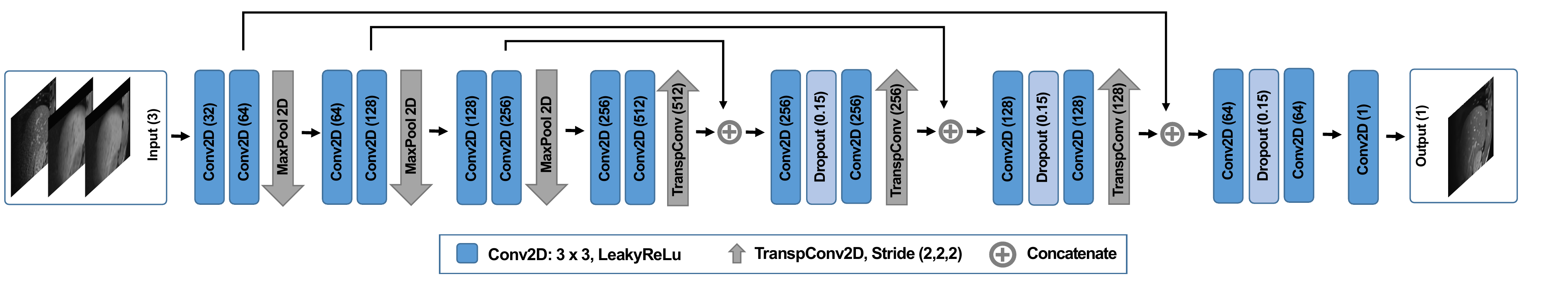}
    \end{center}
    \caption{U-Net variant. Blue boxes are convolutions, grey arrows denote max pooling or up sampling, pluses denote concatenations of feature maps.}
    \label{fig:network}
\end{figure}

\subsection{U-Net CNN Architecture}
\label{sec:architecture}

Our CNN architecture, depicted in Fig. \ref{fig:network}, is a U-Net \cite{ronneberger2015u}.
The input to the network is processed in an encoding and decoding path. The encoding employs four blocks of $3\times3$ convolutional layers, each followed by a leaky rectified linear unit (leaky ReLu) with a slope coefficient of \num{0.1}. The convolutions are padded to keep the size of feature maps and input. The second convolutional layer in each block doubles the number of features, increasing the network's capacity. It is followed by a MaxPooling operation in the first three blocks. The 128$\times$128$\times$3 input to the network is processed by 32 filters in the first convolutional block and results in 512 filters in the latent feature space. 
The decoding reconstructs the image from the latent space. To this end, three transposed convolutional blocks upsample the features, each of which consists of two convolutional layers with a dropout layer in between. At each upsampling, the filter size is halved. At the end, a final $1\times1$ convolution layer outputs the reconstructed image. 
As with the original U-Net architecture, skip connections are used to forward details from the encoding path to the decoding path. In total, the network has about 6,8 million parameters, which get trained by an Adam optimizer \cite{kingma2014}. We implemented the network with Keras \cite{chollet2015}.

\subsection{Training}
\label{sec:training}
\begin{figure}[ht]
    \begin{center}
        \includegraphics[width=.85\textwidth]{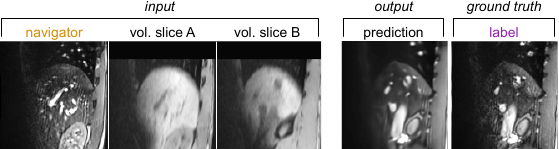}
    \end{center}
    \caption{The network input consists of three channels. The first channel gets a navigator slice that tells which breathing state to predict, i.e., the breathing state that follows the navigator. The second channel gets a static volume slice (vol. slice A) at the navigator position, to act as a still reference to the moving navigator. The third channel gets a static volume slice (vol. slice B) that tells the network at which position to predict the new slice.}
    \label{fig:in_out}
\end{figure}

First, we split the data set into training and validation data (16 subjects) and test data (4 subjects).
We want to predict a 3D volume from a navigator slice. However, for each time point, we do not have ground truth for the whole volume, just one single slice in that volume, which is the data slice following the navigator in an interleaved sequence.
Thus, we trained our U-Net to predict 2D liver slices of a subject given a moving navigator frame and two slices from a static 3D liver volume of that subject. Fig. \ref{fig:in_out} depicts an example for the three channel input, the output and the training label. Vol. slice A is a slice from the static 3D volume and acts as a still reference at navigator position. Vol. slice B is a slice from the static 3D volume at the position for which the network is to predict a liver slice that matches the breathing state of the navigator. This way, the network is able to predict any slice position within the liver and, by that, to slice-wise predict a full 3D liver volume with the adequate breathing state and contrast. To be precise, because the navigator slice is acquired \SI{166}{\ms} before the label data, the network predicts data slices ahead of time.

For training, the navigator slice is taken from the dynamic interleaved sequences (see Fig. \ref{fig:teaser}). The training label (ground truth) is the second image of each pair in an interleaved sequence. The two static volume slices are taken from the static STAR VIBE volume. The first static volume slice is at navigator position, while the second is at data slice position, i.e., at label position. The network can infer the breathing state from the first two input channels, while it can detect and interpret the position from the third input channel. Changing the first input channel changes the breathing state. Changing the third input channel changes the slice position to be predicted. This way, it is possible to slice-wise predict full liver volumes for a dynamic sequence of navigator slices of arbitrary length. 

Before converting to image arrays during the training pipeline, the MRIs were re-sampled using scanner coordinates. Re-sampling was done to harmonize the network input. The slices were re-sampled to 128$\times$128 voxels with a size of $1.8\times\SI{1.8}{\mm}^2$. The STAR VIBE volume was re-sampled to 209$\times$128$\times$128 voxels with a size of $1.8\times1.8\times\SI{1.8}{\mm}^3$.

To facilitate robustness, we augmented the training data in physiological plausible ranges in-plane with random translation of up to $\pm$10 voxel  (\SI{\pm18.18}{\mm}), random rotation of up to \ang{\pm3} and random scaling within ${[}0.8, 1.2{]}$. We evaluate the augmentation in Sec. \ref{sec:experiments_and_results}. 
Furthermore, we whiten image intensities $I$ for each subject using

\begin{equation} 
    \label{eq:norm}
    I_{norm} = \frac{I - \mu}{{\sigma^2}_{adj}},\hspace{30pt} \text{and} \hspace{30pt} {\sigma^2}_{adj} = \max\left({\sigma^2}, \frac{1}{\sqrt{\#voxels}}\right),
\end{equation}

\noindent where $I_{norm}$ are the whitened (normalized) intensities, $\mu$ is the average intensity for all slices of one subject. Likewise, ${\sigma^2}_{adj}$ is the standard deviation, which was adjusted by a reasonable lower bound that depends on the number of voxels $\#voxels$ available for that subject.

We trained a network for each of the 16 subjects. To this end, we split the available samples of each subject into 8,811 training and 180 validation samples (roughly 4 validation samples per slice position). The networks were trained for 200 epochs using the mean squared error (MSE) between prediction and label image as loss function.

\section{Experiments and Results}
\label{sec:experiments_and_results}

We can utilize the network in two ways. First, we can use the network to predict 3D liver MRI in near real-time for any navigator image during interventions. Second, we can use the network to reconstruct a 4D liver MRI from a sequence of navigator slices. In both cases, an input batch is constructed, where each entry of the batch corresponds with a slice position in the reconstructed volume. This allows us to infer all slices for a 3D volume in a single forward pass. After inference, the predicted 2D slices are concatenated to a volume using the scanner coordinates of the STAR VIBE volume. Note that within one batch all inputs have the same navigator slice (first channel), while all third channels show different positions of the static liver volume. This process may be repeated several times to form any 4D breathing sequence. The approach can suffice with \SIrange{2}{24}{\min} of training data (acquisition time), which will be shown later in this section, and the computation time for one 3D prediction is \SI{0.6}{\s}. We compare acquisition and reconstruction times with state of the art methods in table \ref{tab:times}. Note that our method can be combined with the last three methods, which would lower the acquisition times further.  

\begin{figure}[ht]
    \begin{flushleft}
        \includegraphics[width=.9\textwidth]{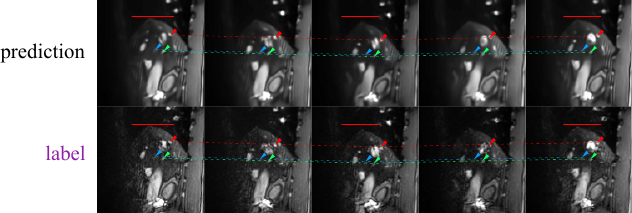}
    \end{flushleft}
    \caption{Sample breathing cycle (prediction and ground truth) for TRE calculation with tracked vessels (arrows) and their traces (dashed lines). Red lines serve as reference for the breathing depth. For compactness, only every second time point is shown. The network input to the first time point is shown in Fig. \ref{fig:in_out}. Slice position is \SI{3.8}{\cm} left of the navigator.}
    \label{fig:breathing_cycle}
\end{figure}

\begin{table}[]
  \begin{center}
\begin{tabular}{lll}
 method & acquisition time &  reconstruction time\\ \hline
 ours & \SIrange{2}{24}{\min} &  \SI{0.6}{\s}\\
 Yuan et al. \cite{yuan2019} & * & \SI{0.615}{\s} \\
 von Siebenthal et al. \cite{siebenthal2007} & \SIrange{15}{60}{\min} & \SI{73}{\s} \\
 Gulamhussene et al. \cite{gulamhussene2020} & \SIrange{15}{60}{\min} & \SI{24}{\s} \\
 Tanner et al. \cite{tanner2014} & \SIrange{9}{12}{\min} & - \\
 Celicanin et al. \cite{celicanin2015} & \num{1/2} & -  \\
 Zhang et al. \cite{zhang2018} & \num{1/4} & \num{1/2} (for sorting approaches)
\end{tabular}
\end{center}
\caption{Comparison of acquisition time (of training or stacking data) and reconstruction time (per time point), if reported. * Method two reconstructs during acquisition.}
\label{tab:times}
\end{table}

\emph{We computed the target registration error (TRE)} of manually tracked vessel cross-sections for all four test subjects. The TRE is a medically crucial metric for accuracy. Vessel cross-sections were tracked within slice positions, which were at different distances to the navigator. Per subject, we chose four slice positions, i.e., four interleaved sequences, at roughly \SI{-3}{\cm}, \SI{-2}{\cm}, \SI{0}{\cm} and \SI{3}{\cm} distance to the navigator (signs indicate left or right of navigator). For each of these 16 interleaved sequences (4 subjects $\times$ 4 slice positions) one random breathing cycle (exhale to exhale, between 8 to 17 time points) was used to track one to six vessel cross-sections (depending on visibility) in the label slice of each time point, i.e., the second slice of a navigator-data pair. By that, we generated a ground truth representing the positions of vessel cross-sections in 16 breathing cycles. We then reconstructed the same 16 breathing cycles (2D sequences) using our CNN approach and manually tracked the same vessel cross-sections again in each prediction. That way, we generated a total of \num{1566} data points in both ground truth and prediction that were used for TRE evaluation. Fig. \ref{fig:breathing_cycle} shows an example breathing cycle and its prediction used for TRE calculation. Note how the network enhances the image quality compared to the label and even predicts vessels correctly that are barely visible in the ground truth label. The input to the prediction of time point one is the same as shown in Fig. \ref{fig:in_out}. The inputs to the other time points look similar to the first, only the navigator changes as it advances in time.
By calculating the mean difference in tracked vessel positions between ground truth and prediction, we computed the TRE per subject and slice position, i.e., interleaved sequence, for a total of 16 calculated TREs shown in table \ref{tab:tre}. All reported TREs are below voxel size, with one exception (S1 column 5). All subjects have a similar over all TRE. The mean TRE for all test subjects is \num{0.66\pm 0.41}\si{voxel} (\num{1.19\pm 0.74}\si{\mm}). One can also see that, in general, the TRE is smaller near the navigator than further away from the navigator.  

\begin{table}[]
  \begin{center}
  \begin{tabular} {clllll}
  \toprule
  \multicolumn{1}{c|}{}  & \multicolumn{5}{c}{slice distance to navigator}             \\ \cline{2-6} 
  \multicolumn{1}{c|}{subject} & \multicolumn{1}{c|}{\SI{3}{\cm}}                       & \multicolumn{1}{c|}{\SI{0}{\cm}}                       & \multicolumn{1}{c|}{\SI{-2}{\cm}}                      & \multicolumn{1}{c|}{\SI{-3}{\cm}}                    & all positions                    \\ \hline  
  \multicolumn{1} {c|}{S1}     & \multicolumn{1} {c|}{\footnotesize\num{0.70\pm 0.38} } & \multicolumn{1} {c|}{\footnotesize\num{0.53\pm 0.37} } & \multicolumn{1} {c|}{\footnotesize\num{0.92\pm 0.62} } & \multicolumn{1}{c|}{\footnotesize\num{1.21\pm 0.65}} & \footnotesize\num{0.84\pm 0.5}   \\
  \multicolumn{1} {c|}{S2}     & \multicolumn{1} {c|}{\footnotesize\num{0.66\pm 0.40} } & \multicolumn{1} {c|}{\footnotesize\num{0.64\pm 0.47} } & \multicolumn{1} {c|}{\footnotesize\num{0.80\pm 0.49} } & \multicolumn{1}{c|}{\footnotesize\num{0.80\pm 0.54}} & \footnotesize\num{0.72\pm 0.47}  \\
  \multicolumn{1} {c|}{S3}     & \multicolumn{1} {c|}{\footnotesize\num{0.44\pm 0.27} } & \multicolumn{1} {c|}{\footnotesize\num{0.45\pm 0.34} } & \multicolumn{1} {c|}{\footnotesize\num{0.60\pm 0.30} } & \multicolumn{1}{c|}{\footnotesize\num{0.94\pm 0.55}} & \footnotesize\num{0.61\pm 0.36}  \\
  \multicolumn{1} {c|}{S4}     & \multicolumn{1} {c|}{\footnotesize\num{0.45\pm 0.27} } & \multicolumn{1} {c|}{\footnotesize\num{0.31\pm 0.21} } & \multicolumn{1} {c|}{\footnotesize\num{0.53\pm 0.38} } & \multicolumn{1}{c|}{\footnotesize\num{0.58\pm 0.33}} & \footnotesize\num{0.47\pm 0.37}   \\ \hline
  \multicolumn{1} {c|}{S1-S4}  & \multicolumn{1} {c|}{\footnotesize\num{0.56\pm 0.33} } & \multicolumn{1} {c|}{\footnotesize\num{0.48\pm 0.35} } & \multicolumn{1} {c|}{\footnotesize\num{0.71\pm 0.45} } & \multicolumn{1}{c|}{\footnotesize\num{0.88\pm 0.52}} & \footnotesize\num{0.66\pm 0.41}    
  \end{tabular}
  \end{center}
  \caption{Presented are TREs for all test subjects. Columns 2-5 show TREs per slice position, the last column shows mean TREs per subject, i.e., over all four slice positions. The last row shows mean TREs over all test subjects. All TREs are given in voxel.}
  \label{tab:tre}
  \end{table}

\begin{figure}[ht]
    \begin{flushleft}
        \includegraphics[width=.9\textwidth]{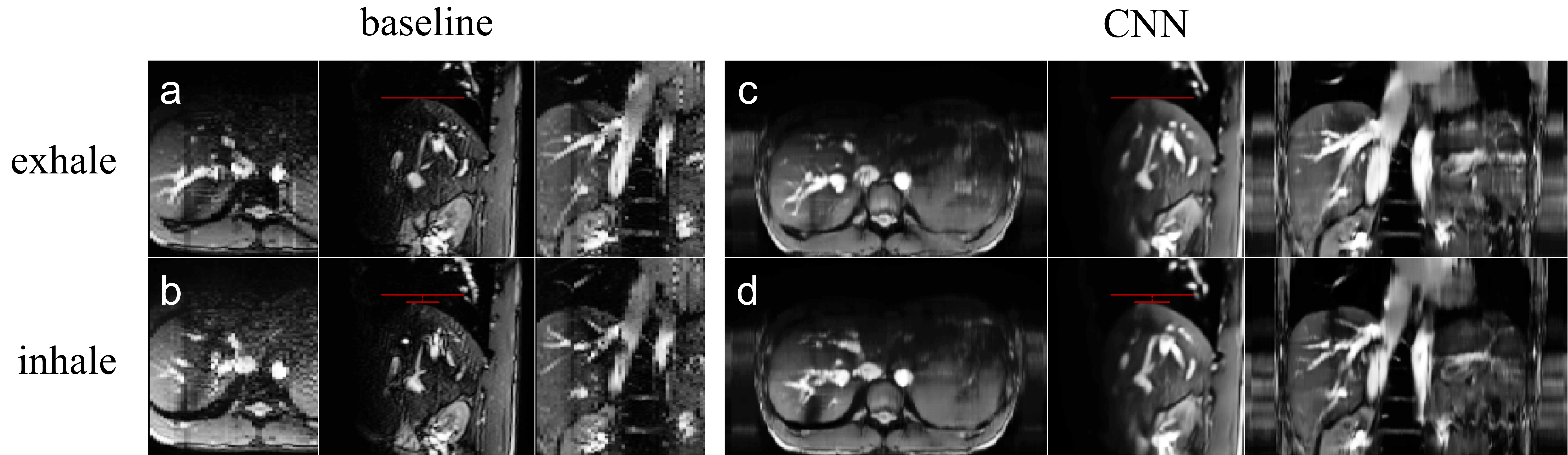}
    \end{flushleft}
    \caption{Example reconstruction of baseline and our CNN-based method, presented as axial, sagittal and coronal slices at identical temporal and spacial position, for an exhale-state (a,c) and inhale-state (b,d). Red lines indicate liver dome position of baseline reconstruction.}
    \label{fig:comparison_tu_cnn}
\end{figure}

\emph{We visually assessed reconstruction results} of our method by analyzing the 4D liver MRIs for all subjects predicted from the reference sequences. We compare our results against a state of the art method. To that end, we used the reconstruction method of Gulamhussene et al. \cite{gulamhussene2020} as baseline. All reconstructions were visually plausible when compared with the baseline reconstruction. In Fig. \ref{fig:comparison_tu_cnn} we present the same subject from the test data set that was used in Fig. \ref{fig:in_out} and \ref{fig:breathing_cycle} to illustrate the reconstruction results. More precisely, the reconstructed end-exhale and end-inhale time points are shown for the baseline and our CNN-based reconstruction. For both the exhale and inhale baseline reconstructions (a,b), we observe that blood vessels and liver boundaries are continuous and smooth in axial and coronal views. Please note that both methods, the baseline reconstruction and our CNN-based method, are performed slice-wise from sagittal slices. Additionally, in our CNN exhale and inhale reconstructions (c,d), the liver dome and vessels are continuous along all view axes. The major vessels are present and smooth. In-plane details are well reconstructed, however some smaller vessels are missing or do not show the correct trajectory in axial orientation, especially in slices further away from the navigator. The breathing depths match excellently between baseline and our method. Note that, to a limited degree, our proposed method is capable of reconstructing regions of the thorax and abdomen left and right that it never saw during training. The baseline method (or any other sorting method) cannot reconstruct these regions. 

\begin{figure}[ht]
    \begin{flushleft}
        \includegraphics[width=.92\textwidth]{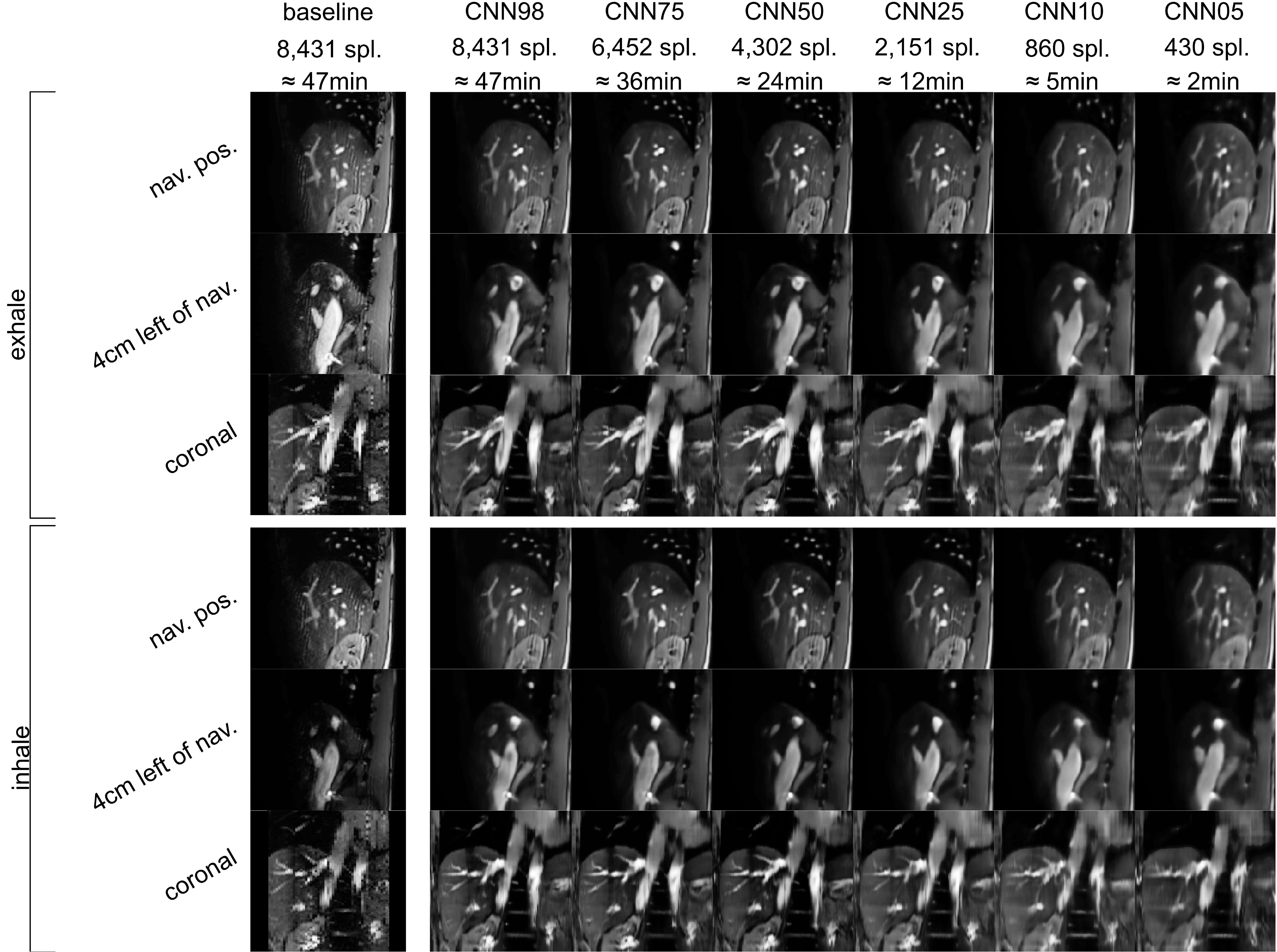}
    \end{flushleft}
    \caption{Reconstruction results depending on training data size. Reconstructed are an inhale and exhale state. The training data size in samples (spl.) is depicted at the top. For each volume reconstruction three slices are presented: two sagittal slices, one at navigator position (nav. pos.), one \SI{4}{\cm} left of the navigator and one coronal slice.}
    \label{fig:less_data}
\end{figure}

\emph{We evaluated the reconstruction quality w.r.t. the training data size.} Fig. \ref{fig:less_data} shows the baseline reconstructions (leftmost column) as well as reconstructions from six networks (other columns) with decreasing amounts of training data (98\% to 5\%). The acquisition times range between \SI{47}{\min} (8431 samples) and \SI{2}{\min} (430 samples). As can be seen, our method is capable of reconstructing full-liver volumes with different breathing states while capturing major and minor vessels. We observe that \SI{2}{\min} of training data yield promising results. Note that with a standard MRI acquisition it would take roughly that time to capture only one 3D volume with comparable quality. We further observe that 50\% of the training data yields nearly as good results as when 98\% of the data are used. 

\begin{figure}[ht]
    \begin{center}
        \includegraphics[width=\textwidth]{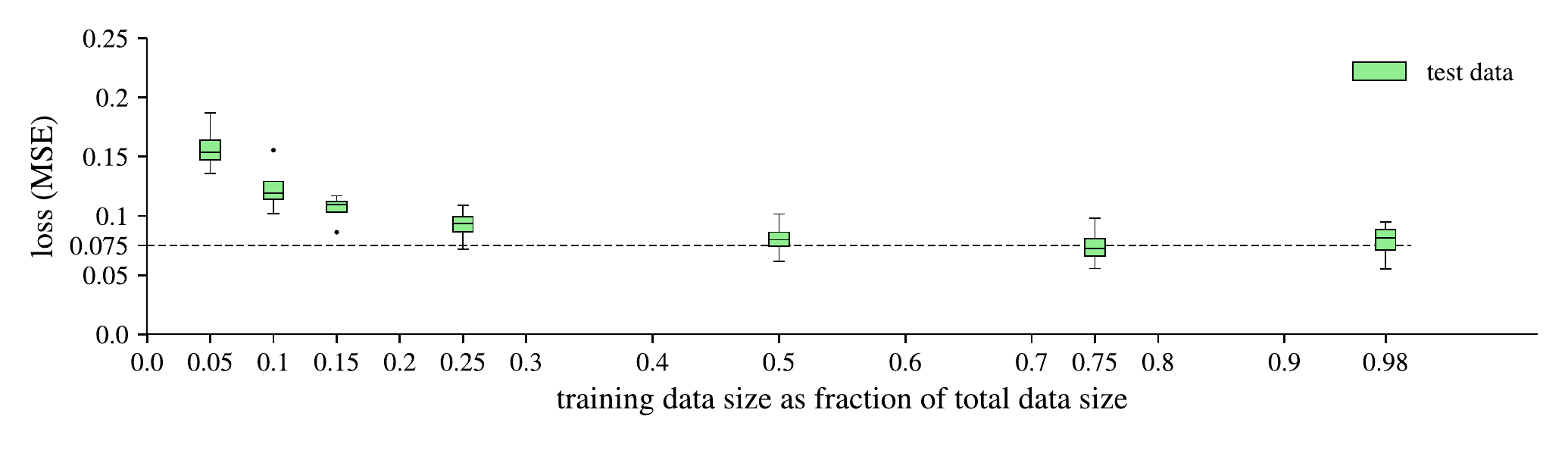}
    \end{center}
    \caption{Test data losses (y-axis) as a function of the amount of available training data (x-axis), starting with 5\% and ending with 98\%.}
    \label{fig:less_data_line}
\end{figure}

Also, in Fig. \ref{fig:less_data_line}, we observe that increasing the training data size beyond 50\% does not improve the loss further, as the latter plateaus at 0.075, indicating that a fraction of the acquisition time would be sufficient for a satisfactory reconstruction.

\begin{figure}[ht]
    \begin{center}
        \includegraphics[width=\textwidth]{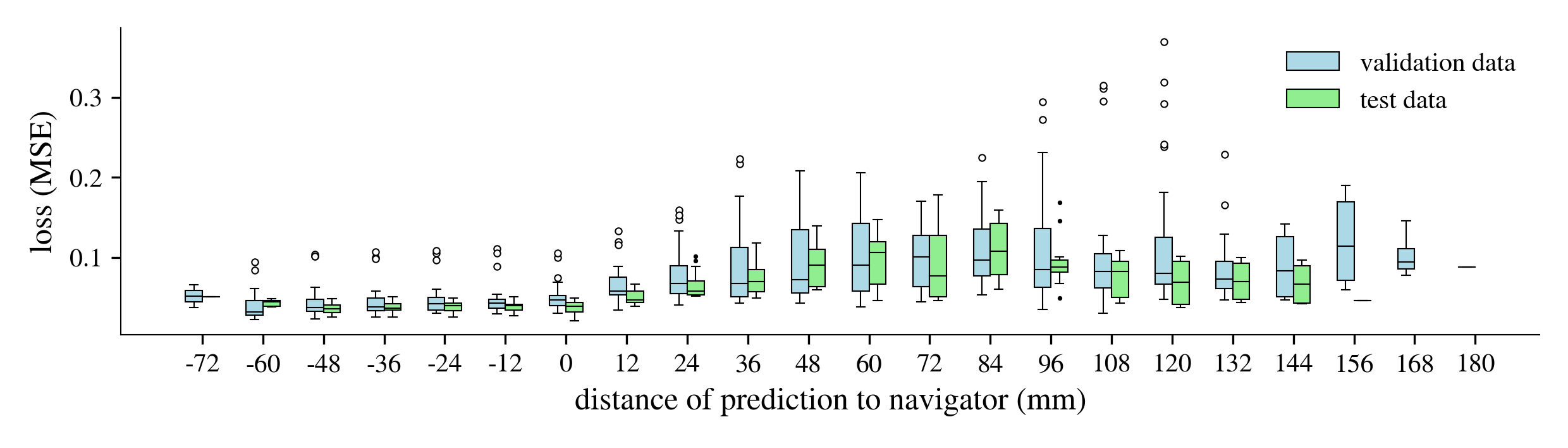}
    \end{center}
    \caption{Loss as function of the distance of prediction to navigator position. Blue box plots are validation data losses (16 subjects). Green box plots are test data losses (4 subjects).}
    \label{fig:val_pos}
\end{figure}

\emph{We evaluated the performance of our method per predicted position.} We used the CNN50 (50\% of the subject data used) models for this analysis. Fig. \ref{fig:val_pos} shows the loss as a function of the distance of the prediction to the navigator slice. Blue and green boxes represent the validation and test data losses respectively. For visualization, the distances were binned into \SI{12}{\mm} bins (3 slice positions per bin). We observe that the test data loss is comparable with validation data loss. Two effects are visible for both data sets. First, our network performs better on the left of the navigator (subjects' right) and worse on the right (subjects' left). Second, our network performs better when being closer to the navigator. This is consistent with the observation made above in the analysis of the TRE. 

\emph{We performed a hyper parameter search} across the 16 subjects of the training and validation data to find the following best settings: learning rate (0.000413), drop out ratio (0.15), data shuffling (true) and batch normalization (false). We also tested the augmentation parameters and found them to improve the reconstruction results for all subjects irrespective of the exact range of any single parameter.

\section{Discussion and Conclusion}
\label{sec:discussion_and_conclusion}

In this work, we present a novel method in which a trained CNN is used to reconstruct high quality 4D full-liver MRI end-to-end in near real time. Our method can solve the presented problem in two distinct ways: First, it can predict a live 3D liver MRI from a live 2D navigator slice in near real-time (\SI{0.6}{\s}) and second, it can reconstruct a 4D liver MRI from a 2D navigator MRI sequence with high temporal resolution (\SI{0.166}{\s}) in retrospection. The reconstruction quality is comparable to the state-of-the-art. Our TRE is well below voxel resolution with \num{0.66\pm 0.41}\si{voxel} (\num{1.19\pm 0.74}\si{\mm}), which is medically sufficient. 

We showed that, with our method, acquisition times could be halved from \SI{47}{\min} (\num{8431} training samples) to \SI{24}{\min} (\num{4302} samples) without losing reconstruction quality. They can even be reduced to \SI{2}{\min} (\num{430} samples) while losing some image quality. Some of the earlier proposed methods report acquisition times between \num{9} and \SI{60}{\min}, while others report acquisition time reductions between \num{1 / 4} and \num{1 / 2}. Because our new method complements these methods, it can be used in conjunction to multiply the reduction effects. Thus, a combined acquisition time reduction of up to \num{3 / 4} without loss of reconstruction quality seems achievable. In practice, to reconstruct breathing sequences of arbitrary length, this would mean acquisition times of around \SI{6}{\min}, which is a reasonable time in clinical practice. Overall, our work shows promising results which could motivate further research in this direction. We believe our method shows a way for predicted true real-time 4D MRI techniques and provides a solution to reduce the acquisition time and effort for retrospective reconstruction approaches. 

Summarily, the three key strengths of our method, from a medical point of view, are high reconstruction accuracy, high image quality and resolution, and high speed in both acquisition and reconstruction. The main contribution of this paper is to prove the possibility of applying end-to-end deep learning on the problem of 4D MRI reconstruction to achieve these key strengths.

In the following, we discuss the limitations of our method and motivate future work. Because our network is 2D, it cannot acquire full knowledge of 3D relations between navigator and data slices. The further away the data slice is from the navigator, the looser the 3D relations become, and the poorer the reconstruction quality end up. To mitigate this effect, one could potentially divide the volume into distance ranges and train one network for each range, thus reinforcing knowledge for 3D relations over larger distances. We expect that an ensemble of such networks will provide a considerable gain in quality for a fixed level of training data or constant quality for less training data. 

Additionally, in our method, one model is trained for each subject. In future work, we want to investigate the possibility of having only one model that abstracts not only beyond seen breathing states, but also beyond seen subjects, or adapts quickly to new subjects. We believe that this is achievable using transfer learning strategies which, in turn, will further reduce the amount of necessary training data.

\bibliography{literature}{}
\bibliographystyle{plain}

\end{document}